\documentclass[fleqn,10pt]{olplainarticle}

\usepackage[T1]{fontenc}
\usepackage[utf8]{inputenc}
\usepackage{microtype}
\usepackage{lmodern}
\usepackage{comment}
\usepackage{mathtools}
\usepackage{amssymb}
\usepackage{bbm}
\usepackage{amsthm}
\usepackage{diffcoeff}
\usepackage{subcaption}
\usepackage[]{graphicx}
\usepackage{float}
\usepackage{siunitx}
\usepackage[colorinlistoftodos]{todonotes}
\usepackage[backend=biber,
    citestyle=authoryear-icomp,
    natbib=true,
    maxbibnames=99,
    maxcitenames=2,
    giveninits=true,
    terseinits=true,
    uniquelist=false]{biblatex}
\usepackage{csquotes}

\usepackage{hyperref}
\usepackage{xcolor}
\definecolor{rsrs}{rgb}{0.0, 0.2, 0.480}
\hypersetup{
    colorlinks,
    linkcolor={rsrs},
    citecolor={rsrs},
    urlcolor={rsrs}
}
\usepackage[noabbrev]{cleveref}

\DefineBibliographyStrings{english}{%
    andothers = {\emph{et al}}
}
\DeclareNameAlias{author}{last-first}
\DeclareFieldFormat*{title}{#1}
\renewbibmacro{in:}{}

\citetrackerfalse
\DefineBibliographyStrings{english}{%
  mathesis = {MSc thesis},
}

\title{Probabilistic dose prediction using mixture density networks for automated radiation therapy treatment planning}

\author[1]{Viktor Nilsson$^*$}
\author[1]{Hanna Gruselius\thanks{These authors contributed equally to this work.}}
\author[1, 2]{Tianfang Zhang}
\author[3]{Geert De Kerf}
\author[3, 4]{Michaël Claessens}
\affil[1]{ RaySearch Laboratories, Stockholm, Sweden}
\affil[2]{ Mathematical Statistics, Department of Mathematics, KTH Royal Institute of Technology, Stockholm, Sweden}
\affil[3]{ Iridium Cancer Network, Antwerp, Belgium}
\affil[4]{ Department of Radiation Oncology, Faculty of Medicine and Health Sciences, University of Antwerp, Antwerp, Belgium}


\begin{abstract}
We demonstrate the application of mixture density networks (MDNs) in the context of automated radiation therapy treatment planning. It is shown that an MDN can produce good predictions of dose distributions as well as reflect uncertain decision making associated with inherently conflicting clinical tradeoffs, in contrast to deterministic methods previously investigated in literature. A two-component Gaussian MDN is trained on a set of treatment plans for postoperative prostate patients with varying extents to which rectum dose sparing was prioritized over target coverage. Examination on a test set of patients shows that the predicted modes follow their respective ground truths well both spatially and in terms of their dose--volume histograms. A special dose mimicking method based on the MDN output is used to produce deliverable plans and thereby showcase the usability of voxel-wise predictive densities. Thus, this type of MDN may serve to support clinicians in managing clinical tradeoffs and has the potential to improve quality of plans produced by an automated treatment planning pipeline.

\end{abstract}

\begin{document}

\flushbottom
\maketitle
\thispagestyle{empty}

\section*{Introduction}
Radiation therapy is a cornerstone treatment used as single or combined treatment for more than 40 percent of cancer patients  \parencite{borras2016many}.
Previous and emerging innovations in imaging and treatment techniques have contributed to a more accurate radiation administration to the tumor while simultaneously sparing organs at risk.
Despite these improvements, different processes of the radiation therapy workflow are still susceptible for variations.
Important sources of uncertainty include heterogeneity in the delineation of regions of interest (ROIs) \parencite{brouwer20123d, van2019interobserver} and differences in preferences associated with conflicting clinical tradeoffs for a particular treatment site \parencite{nelms2012variation}.
In particular, the radiation therapy treatment planning (RTTP) process requires interaction between dosimetrists, medical physicists and radiation oncologists to find the best treatment plan for each patient.
The creation of a deliverable plan typically involves repeatedly solving a large-scale optimization problem with iteratively updated parameters, which makes it a time-consuming and labor-intensive step \parencite{craft2012improved}. 
Moreover, the process is characterized by a user-variability both intra- and inter-institutional due to subjectivity in the management of tradeoffs between planning objectives and institution-specific guidelines \parencite{nelms2012variation}.
Therefore, efforts have been made to develop methods that can automate the RTTP process to reduce the time required for human intervention and the variation between dosimetrists/physicists to improve overall plan quality and consistency.



In this context, knowledge-based planning (KBP) could remedy the differences coming from subjectivity and has the potential to homogenize the RTTP process  \parencite{berry2016interobserver}. \textcite{ge2019knowledge} divide current methods in KBP in two categories: (1) case- and atlas-based methods and statistical modeling and (2) machine learning (ML) methods.
In the former category, a reduced plan quality variation is achieved by seeking similar patients in a cohort of available ones, and then leveraging the knowledge from the previous case to a new patient. 
In the latter type, a model is trained on a set of previously treated patients, which is then used to predict some dose-related quantity for a new patient that is of assistance when creating a deliverable plan. 
Examples of such quantities are beam angles, dose--volume histograms (DVHs), voxel-wise dose and mean region doses \parencite{ge2019knowledge}. Voxel-wise dose prediction methods have been created using a variety of ML-based approaches. 
\textcite{u-net-dose} validate that a neural network (NN) of U-net type can be applied for voxel-wise dose prediction based on patient contours for prostate intensity-modulated radiation therapy, and \textcite{campbell2017neural} use an NN to predict dose for pancreatic stereotactic body radiotherapy (SBRT). 
Further developments of NNs for KBP have been achieved by \textcite{kearney2018dosenet} and \textcite{shiraishi2016knowledge}, where a fully convolutional three-dimensional NN is created for dose prediction in prostate SBRT. 
Also, generative approaches have been investigated in \textcite{mahmood2018automated}, \textcite{murakami2020fully} and \textcite{babier2020knowledge}, where voxel-wise dose is predicted by generative adversarial networks (GANs). 
Other works have used more classical ML algorithms, e.g. random forest--based methods to predict voxel doses \parencite{mcintosh2016voxel, mcintosh2017fully} and the application of flow and shape models for spine SBRT \parencite{liu2015active}. 

Common for previous NN-based approaches is that a single dose per voxel is outputted by the ML algorithm---the dose of the predicted optimal treatment plan for the patient. However, for complex cases, inter-planner variations with KBP will remain \parencite{KUBO2019132varplanner}. 
Hence, an optimal prediction could consist of the dose distribution that best mimics the behavior of previously seen data, while there is still a possibility to incorporate other preferences by steering the ML algorithm during training. 
This type of KBP can remedy the variability brought by subjectivity of a planner in the RTTP process. 
However, since a single dose value is predicted in each voxel, it will not be able to reflect more than one specific protocol and preference in the model. 
Consequently, if there are potentially conflicting tradeoffs inherent in the treatment site, where one would possibly have different protocols or objectives in place (e.g. complex cases), no such concerns can be reflected in the same ML model. 

The present work is an extension of previous methods by incorporating meaningful variations in the RTTP problem in the output of ML algorithm, in particular demonstrating the variability caused by prioritizing different clinical tradeoffs. This is achieved by using a mixture density network to predict a possibly multimodal probability density function of dose in each voxel. This can add to KBP not only a more homogenized manner of RTTP, but also a way of automatically grasping the viable treatments based on clinically relevant variability, which could be of use as a decision support mechanism in clinical practice. Moreover, an integral part of automated RTTP by use of ML is the optimization to create a deliverable plan from a predicted dose distribution subject to physical delivery constraints, commonly referred to as dose mimicking \parencite{mcintosh2017fully}. While dose mimicking can be done in a range of manners, in this paper, we present the idea of incorporating the predictive probability distributions into the objective function as an application of the implied uncertainties. Among other applications, it is also possible to use the voxel-wise probability densities to post-process the predicted dose distribution to have the desired DVHs in different regions of interest \parencite{mcintosh2017fully}, which constitutes a comprehensive way of articulating additional preferences. Hence, a multimodal voxel dose distribution can bring efficiency and improvements to a range of RTTP applications. 

\section*{Methods and materials}

\subsection*{Model setup}
\begin{figure}[h]
        \centering
        \includegraphics[width=1\linewidth]{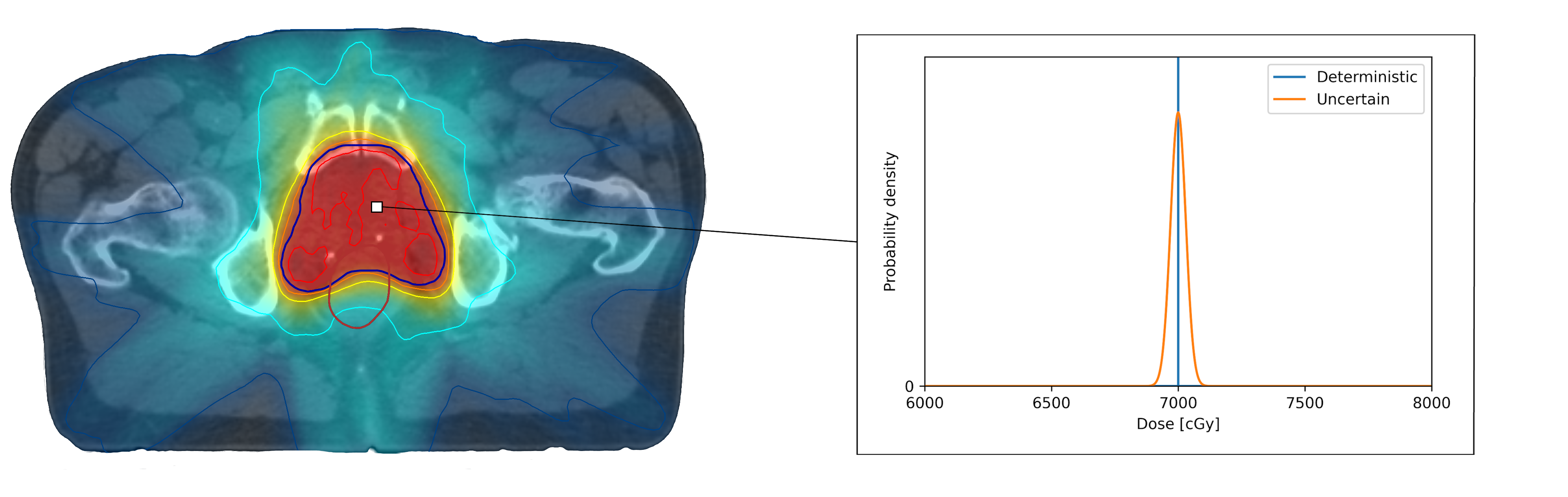}
        \caption{Visualization of a probability distribution of dose versus a deterministic dose for a voxel in the target area.}
        \label{fig:distribution-explanation}
    \end{figure}

    In order to reflect a range of possible predictions with an ML model, its output must comprise more than a deterministic output, such as a single dose value per voxel. 
    A more informative output would instead be a probability distribution per voxel, reflecting the range of possible outcomes of dose values and their respective likelihoods given the planning standards learned from the training data.
    In \cref{fig:distribution-explanation}, an example of such a probability distribution for a voxel in the target area is depicted.  
    
    More formally, let $\{(x_n, d_n)\}_{n=1}^N$ be the training set consisting of independent random variable pairs of contoured images $x_n$ and corresponding dose distributions $d_n$. Given this and the current patient image $x = (x^v)_v$ as a vector over voxels $v$, the task is to obtain for each $v$ the probability distribution of the associated dose $d^v$, i.e.
    \begin{equation}
        \label{margpred}
        p(d^v \mid x, \{(x_n, d_n)\}_{n=1}^N).
    \end{equation}
    Note that this is the marginal predictive distribution for one particular voxel $v$. Although one could, in principle, try to model the joint predictive distribution $p(d \mid x, \{(x_n, d_n)\}_{n=1}^N)$, it is sufficient for the purposes of this study to obtain a collection of marginal distributions over all voxels in the patient volume. 
    
    \subsection*{Mixture density networks}
    
    As mentioned in the introduction, deep learning has been applied to the problem of voxel-wise deterministic dose prediction, e.g. using  U-nets.
    A novel idea is to extend the method presented in \textcite{u-net-dose} around U-net prediction to instead output an inferred predictive distribution $p_{\theta}(d^v \mid x)$, where $\theta$ denotes the network weights. While there are several approaches to making NNs predict stochastic outputs, such as by sampling from a latent variable space \parencite{Goodfellow2014, kingma2013autoencoding}, we shall opt for a discriminative approach where the marginal predictive distributions belong to some parametric family and where the NN outputs are the corresponding parameters. 
    The family of Gaussian mixture models has the favorable property of being dense in the space of probability distributions \parencite{approximation-convolution-defined-mixture}---in other words, it is a parametric family that can approximate any distribution arbitrarily well.
    A one-dimensional Gaussian mixture model with $C$ components is a probability distribution with density function
    \begin{equation}
        \sum_{c=1}^C \alpha_c \operatorname{\mathcal{N}}(\cdot \mid \mu_c, \sigma_c^2),
    \end{equation}
    where $\operatorname{\mathcal{N}}(\cdot \mid \mu_c, \sigma_c^2)$ is the density function of the normal distribution with mean $\mu_c$ and variance $\sigma_c^2$ and where the class probabilities $\{\alpha_c\}_{c=1}^C$ are nonnegative and sum to unity.
    
    Thus, a $C$-component Gaussian mixture model requires $3C$ parameters (alternatively, $3C - 1$, as one is implicit from the constraint that class probabilities sum to unity), which will be outputted by the NN. Such an NN is known as a mixture density network \parencite{Bishop1994MixtureDN}. More precisely, three functions $\alpha_{\theta}$, $\mu_{\theta}$ and $\sigma_{\theta}^2$ will be trained in the network, derived from the network weights $\theta$. $\alpha_{\theta}$ maps each input image $x = (x^v)_v$ to a matrix with $C$ rows, where the output $\alpha_{\theta}(x)_c$ in row $c$ is the vector of the corresponding class probabilities (and similarly for $\mu_{\theta}$ and $\sigma_{\theta}^2$). The inferred voxel-wise predictive likelihood can thus be written as
    \begin{equation}
    \label{voxelwisepredictive}
        p_{\theta}(d^v \mid x) = \sum_{c=1}^C \alpha_{\theta}(x)_c^v \operatorname{\mathcal{N}}\!\left( d^v \mid \mu_{\theta}(x)_c^v, \sigma_{\theta}^2(x)_c^v \right).
    \end{equation}
    
\begin{figure}[h]
        \centering
        \includegraphics[width=0.9\linewidth]{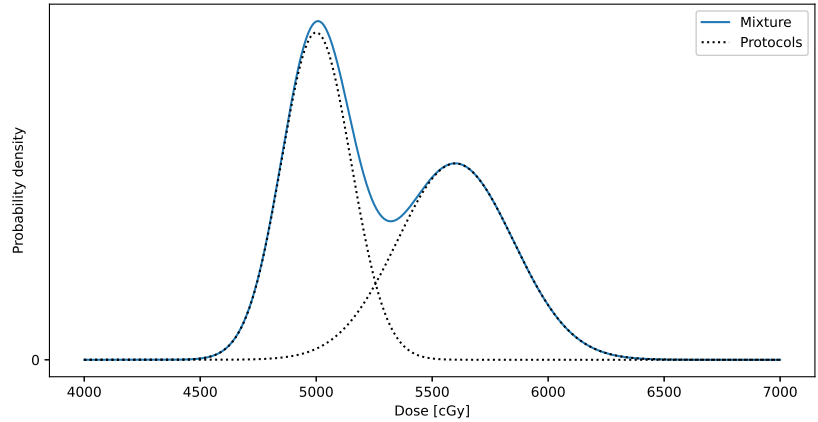}
        \caption{Visualization of how a probability distribution of dose could look like in a voxel, where there are two protocols prescribing different amount of dose to the area. The distribution is a mixture of the two distributions coming from each protocol.}
        \label{fig:distribution-explanation_2}
    \end{figure}

    This type of model is able to reflect any number of inherent distributional components.
    Consider a dose inference problem where one knows there is a tradeoff prevalent in the data stemming from two potential treatment protocols $T_1$, $T_2$.
    An MDN with two components can reflect this behavior by predicting $\mu_1$, $\mu_2$ to match the average behavior of each protocol, and $\sigma_1$, $\sigma_2$ to reflect the internal uncertainty of each prediction.
    In \cref{fig:distribution-explanation_2}, one can see a visualization of how such a predictive density could be in a voxel where this twofold dose preference is present. 
    
\subsection*{Data}
    In order to carry out experiments demonstrating the sought variability for a site, a pertinent clinical tradeoff is needed as a basis for the work. 
    Studies have focused on analyzing the radiobiological effect of restricting the dose to the rectum for prostate patients, possibly at the cost of sacrificing target coverage, which is a clinical tradeoff that is central in prostate RTTP \parencite{mavroidis2017radiobiological}. 
    In this work we will use the dose variation in the rectum area to study the ML method's capability to identify and mimic these tradeoffs in its prediction.
    
    The data used in this work originates from Iridium Cancer Network (Antwerp, Belgium) and consists of prostate cancer patients having undergone a prostatectomy prior to radiation therapy. The dataset contains both patients with and without pelvic nodes included in target volume. All organs were contoured according to the RTOG \parencite{GAY2012e353rtog} and ACROP \parencite{SALEMBIER201849acrop} guidelines. 
    Patients were treated in 35 fractions with a prescribed dose of 56 Gy to the seminal vesicles and eventually also to the pelvic nodes. The prostate bed was treated with a simultaneous integrated boost of 70 Gy. All plans were delivered using two $360^{\circ}$ volumetric modulated arc therapy (VMAT) beams. The clinical goals used during the planning process are summarized in table \ref{tab:clinical_goals}.
    
    \begin{table}[H]
        \centering
                \caption{\label{tab:clinical_goals}Vital clinical goals in the Iridium postoperative prostate protocol.}
        \begin{tabular}{l|r|c}
            Region & Volume $[\%]$ & Dose [Gy]\\\hline
            PTV Prostate & 98 & $\geq 66.5$ \\
                     & 0 & $\leq 74.9$ \\
            PTV Seminal Vesicles & 98 & $\geq 53.2$ \\
            Rectum & 10 & $\leq 65$ \\
                   & 5 & $\leq 70$ \\
            Bladder& 10 & $\leq 70$
        \end{tabular}
    \end{table}
    
\begin{figure}[h]
        \centering
        \includegraphics[width=1\linewidth]{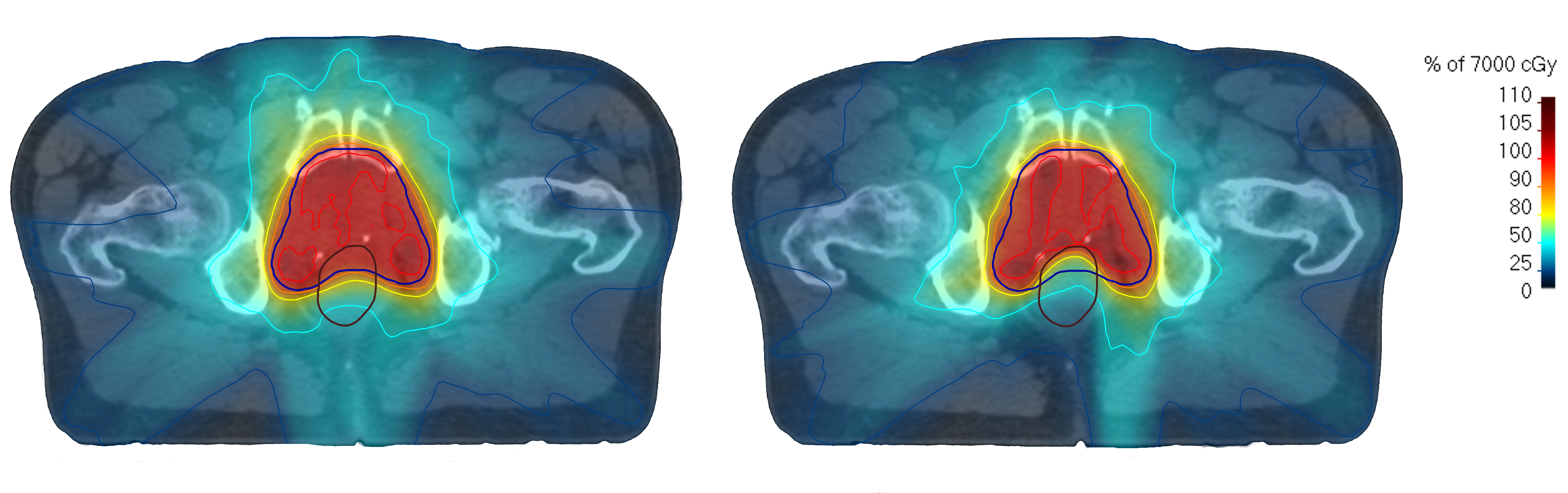}
        \caption{Example of doses used for training. The slice of dose to the left represents the target coverage plan $P_1$ and the right one the the spare rectum plan $P_2$. The slices are from the same patient and position on the longitudinal axis.}
        \label{fig:training-data}
    \end{figure}

    In this work, the training data consisted of a pair $P_1$, $P_2$ of treatment plans created for each patient, where $P_2$ has a dose distribution that spares the rectum to a higher extent than $P_1$, in general at the cost of sacrificing target coverage. In \cref{fig:training-data}, one can see the dose distributions for each such plan in a transverse cross section for a representative patient.
    The training data was generated by application of an ML-based dose prediction and dose mimicking optimization using RayStation 9A (RaySearch Laboratories, Stockholm, Sweden).
    The ML algorithm used to predict the doses was trained on approximately 100 clinical plans and is based on the method provided in \textcite{mcintosh2016voxel}.  
    The model used to create the training data was configured to predict one plan $P_1$ that prioritizes target coverage and another plan $P_2$ that prioritizes sparing of the rectum area. 
    By applying this planning process for 16 patients from the Iridium dataset, 32 plans were created for the training dataset. The objective of using this dual input of plans for each patient is that one can easily validate that this twofold dose preference in the rectum is reflected in new MDN predictions. 
    Rather than creating two clinically acceptable plans, $P_1$ and $P_2$ just reflect two different planning protocols. 
    
\subsection*{Implementation}
    The architecture of the neural network used in this work is a U-net \parencite{ronneberger}. 
    U-nets are convolutional neural networks with dimension $128 \times 176$ in both the input and output layers, and possibly differing number of channels.
    For this study, we restrict our attention to predicting dose in each transversal slice separately. 
    As mentioned in the introduction, this type of approach has been used for dose prediction successfully. 
    
    The input to the U-net could be a CT image with radiodensities and/or contoured ROIs. 
    As the latter alone has proven to be successful for the task, the network built for this work uses only the ROI contours. 
    The ROIs included are the External (the full body), PTV Prostate, PTV Seminal Vesicles, Rectum and Bladder. Hence, the input layer has five channels.
    The simplest representation of the input ROIs are their binary encoding in the three-dimensional patient volume, but since the prediction is based on transversal slices, proximity information of ROIs in the craniocaudal direction would be lost.
    To account for this, a Euclidean distance transform is applied to each ROI before separating the slices.
    That is, letting $d_{\mathrm{Eucl}}$ denote spatial distance, the Euclidean distance transform $T_R$ with respect to a ROI $R$ is a map from voxel to scalar, given by 
    \begin{equation}
        T(v) = \min_{v' \in R} d_{\mathrm{Eucl}}(v, v').
    \end{equation}
    In this work, and augmented version $\hat{T}_R$ given by $\hat{T}_R(v) = T_R(v) - T_{R^{\mathrm{c}}}(v)$, $R^{\mathrm{c}}$ denoting the complement of $R$, is used to provide more local information when $v \in R$.
    
\begin{figure}[h]
        \centering
        \includegraphics[width=.7\textwidth]{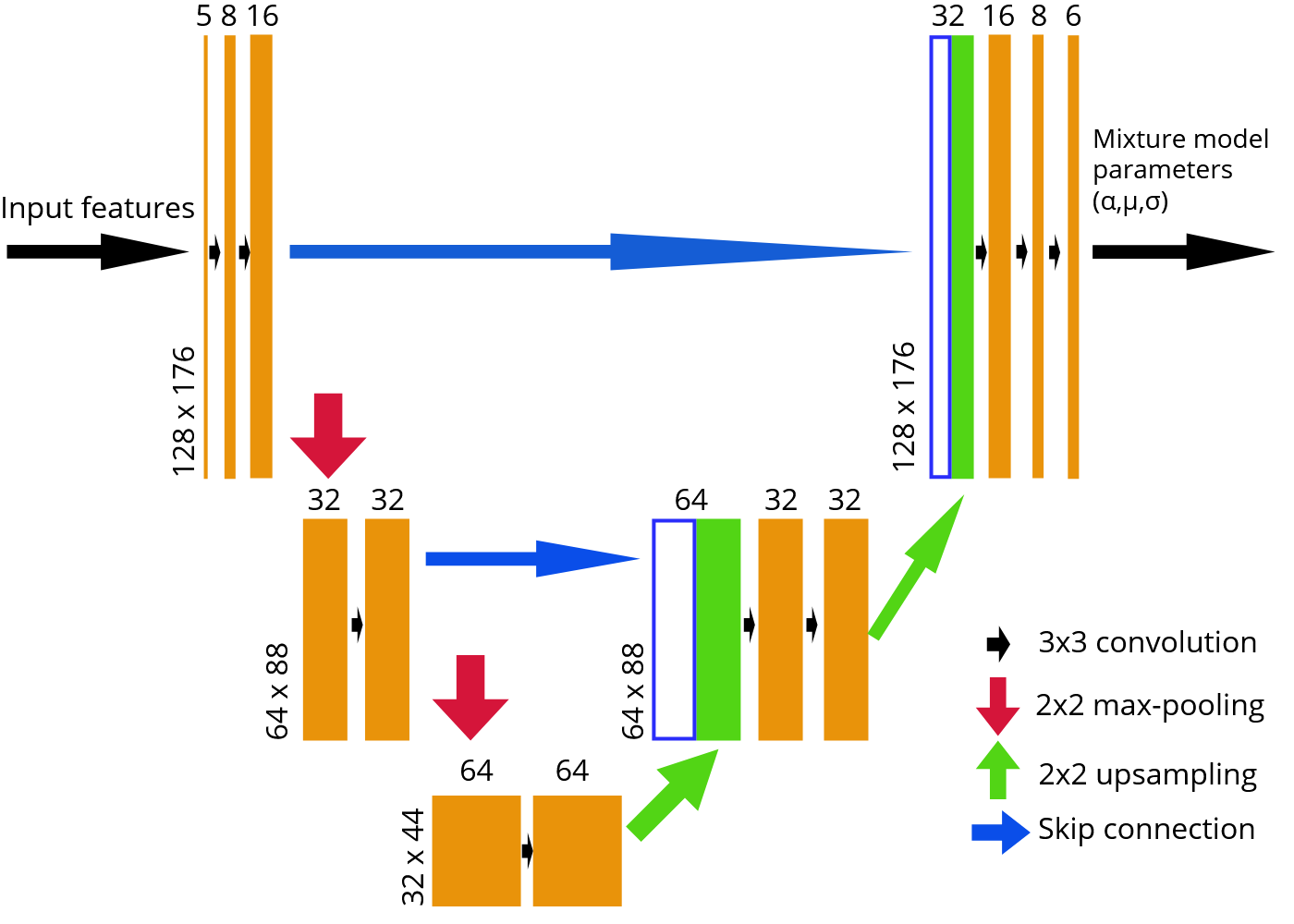}
        \caption{Architecture of the U-net. Each bar represents a layer and the number on top of each the number of channels.}
        \label{fig:unet}
    \end{figure}

    The output of the network is the distributional parameters in each voxel. In this work attention is restricted to a bimodal Gaussian mixture, as the tradeoff we are focusing on comes from two protocols $P_1$, $P_2$, i.e. $C=2$. Thus, the output should have $3C = 6$ channels.
    U-nets have a contractive phase, followed by an expansive phase. 
    During the contractive phase, the image is successively downsampled using max-pooling.
    As the image shrinks, the number of channels is simultaneously increased, which allows higher-level contextual information to be represented. 
    In the expansive phase, this is propagated up to the original resolution using upsampling.
    Fine-grained information may be lost in the contractive part of the network.
    Therefore, there are ``skip connections'' between the phases where detailed input from each step in the downsampling is propagated to the upsamling part. For an illustration of the complete architecture used, see \cref{fig:unet}.
    
    A loss function that reflects the NN's performance in this setting is the negative log-likelihood of the data, which measures how probabilistically congruent our data is to the proposed distributions in each voxel. However, since voxels are not equally critical to the plan, we use a loss function where local negative log-likelihoods $-\log p_{\theta}(d^v \mid x)$, known from Equation~\ref{voxelwisepredictive}, have weights $w^v$ reflecting their relative importance. Additionally, to avoid mode collapse---i.e., making sure that the class probabilities do not vanish---we use a regularization penalty 
    \begin{equation}
        b^v(\theta; x) = \operatorname{max} \{(\alpha_0 - \alpha_{\theta}(x)_c^v)_+\}_{c=1}^C
    \end{equation}
    for each voxel, $\alpha_0$ being some threshold constant. The loss function $L(\theta; x, d)$ associated with the pair $(x, d)$, which in this case are both slice-specific, can thus be written as
    \begin{equation}
        L(\theta; x, d) = \sum_v \left(-w^v \log p(d^v \mid x) + \lambda b^v(\theta; x) \right),
    \end{equation}
    where $\lambda$ is some scaling factor. The constants $\{w^v\}_v$, $\alpha_0$ and $\lambda$ can be seen as hyperparameters of the model and have been configured by hand. For our computational study, we used
    
    \begin{equation}
        w^v = \begin{cases}
            0.05 & \text{if $v$ is in air,} \\
            5 & \text{if $v$ is exclusively in PTV or rectum,} \\
            10 & \text{if $v$ is in PTV and rectum,} \\
            1 & \text{otherwise,}
        \end{cases}
        \text{ }
       \quad \alpha_0 = 0.2, \quad \lambda = 20.
    \end{equation}
    
    Experiments are implemented using TensorFlow 2.2 and TensorFlow Probability 0.10. The model is trained using Adam \parencite{adam} with standard parameters. 
    The final model is selected as the one with the lowest validation loss during training and evaluated on an untouched holdout data set.
    A train/validation/holdout ratio of 50/25/25 is used---specifically, there are 16 patients in the training set, 8 in the validation set and 9 in the holdout set.
    
\subsection*{Dose mimicking}

To demonstrate the usability of predictive probability distributions such as those outputted by the MDN, we use a specially developed dose mimicking method to create deliverable treatment plans according to the predictions. Since the mixtures comprise two components, two plans may be created for each patient---one where the dose should mimic $\mu_1$ and one for $\mu_2$. For each $v$, let $F^v$ be the cumulative distribution function associated with one chosen mixture component of the predictive density $p(d^v \mid x)$, and let $t^v$ be such that $t^v = 0$ if the dose to voxel $v$ is to be minimized and $t^v = 1$ otherwise. In particular, when $v$ belongs to both the PTV and the rectum, we use $t^v = 1$ if mimicking $\mu_1$ and $t^v = 0$ if mimicking $\mu_2$. The objective function $\psi_{\mathrm{Spat}}$ on the spatial dose distribution is written as sum of cross-entropy losses over all voxels in the patient volume---that is,
\begin{equation}
    \psi_{\mathrm{Spat}}(d) = -\sum_v r^v \Big( t^v \log F^v(d^v) + (1 - t^v) \log(1 - F^v(d^v)) \Big),
\end{equation}
where $r^v$ is the voxel volume of $v$ relative to the whole patient volume, satisfying $\sum_v r^v = 1$. To ensure homogeneity in target doses and, in general, that the deliverable dose follows the DVHs of the prediction closely, we add for each ROI $R$ in the model a so called reference-DVH function $\psi_{\mathrm{DVH}, R}$ with $\mu_1$ or $\mu_2$ as reference dose. Their definitions are described in detail in \textcite{albin}. 

The optimizations were performed using direct machine parameter optimization with the same beam settings as used in the original data. Let $\eta$ denote the optimization variables with feasible set $\mathcal{E}$, where the total dose $d$ is determined by some dose deposition mapping $d = d(\eta)$ according to, for example, \textcite{unkelbach}. The optimization problem may then be written as
\begin{equation}
\label{opt}
    \underset{\eta \in \mathcal{E}}{\operatorname{minimize}} \quad w_{\mathrm{Spat}}\psi_{\mathrm{Spat}}(d(\eta)) + \sum_R w_{\mathrm{DVH}, R} \psi_{\mathrm{DVH}, R}(d(\eta)),
\end{equation}
$w_{\mathrm{Spat}}$, $w_{\mathrm{DVH}}$ being weights. For our computational study, we used $w_{\mathrm{Spat}} = 10^{-4}$ and $w_{\mathrm{DVH}, R} = 10^2$ for organs at risk and $w_{\mathrm{DVH}, R} = 10^3$ for targets. The problem above was solved using RayStation's native sequential quadratic programming solver, and final doses were calculated using a collapsed cone algorithm. 

\section*{Results}

The MDN trained for 100 epochs, although after around 50 epochs the final model is typically selected due to incipient overfitting.
After training the network, its performance is assessed through some qualitative inspections.
As it is difficult in general to evaluate fitted predictive distributions, the focus of this section is to provide some key visualizations of the model behavior compared to ground truth counterpart.

\begin{figure}[h]
    \centering
    \includegraphics[width=1\linewidth]{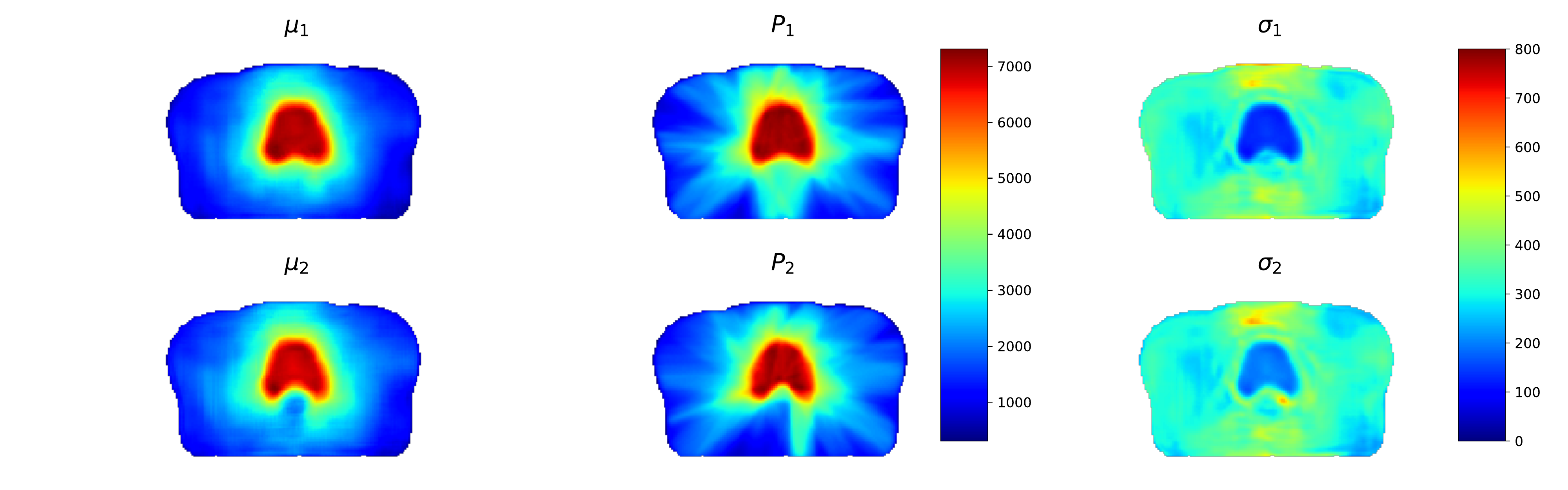} 
    \caption{In left column: expected dose value $\mu_1$, $\mu_2$ [cGy] in a transversal slice from a validation patient. In middle column: dose values in ground truth plans in [cGy] and corresponding slice. In right column:
     parameters $\sigma_1$, $\sigma_2$ in [cGy] the same slice.}
    \label{fig:doses-and-sigma}
\end{figure}

First of all, one would like to see that the two data modes $P_1$, $P_2$ have indeed been separated.
Thus, a central transversal slice from a patient is firstly studied in \cref{fig:doses-and-sigma}, where the two modes $\mu_1$ and $\mu_2$ are plotted.
We also show the predicted standard deviation in each mode.
For reference, the ground truth dose for each data mode $P_1$ and $P_2$ is plotted.

\begin{figure}[h]
    \centering
    \includegraphics[width=0.5\linewidth]{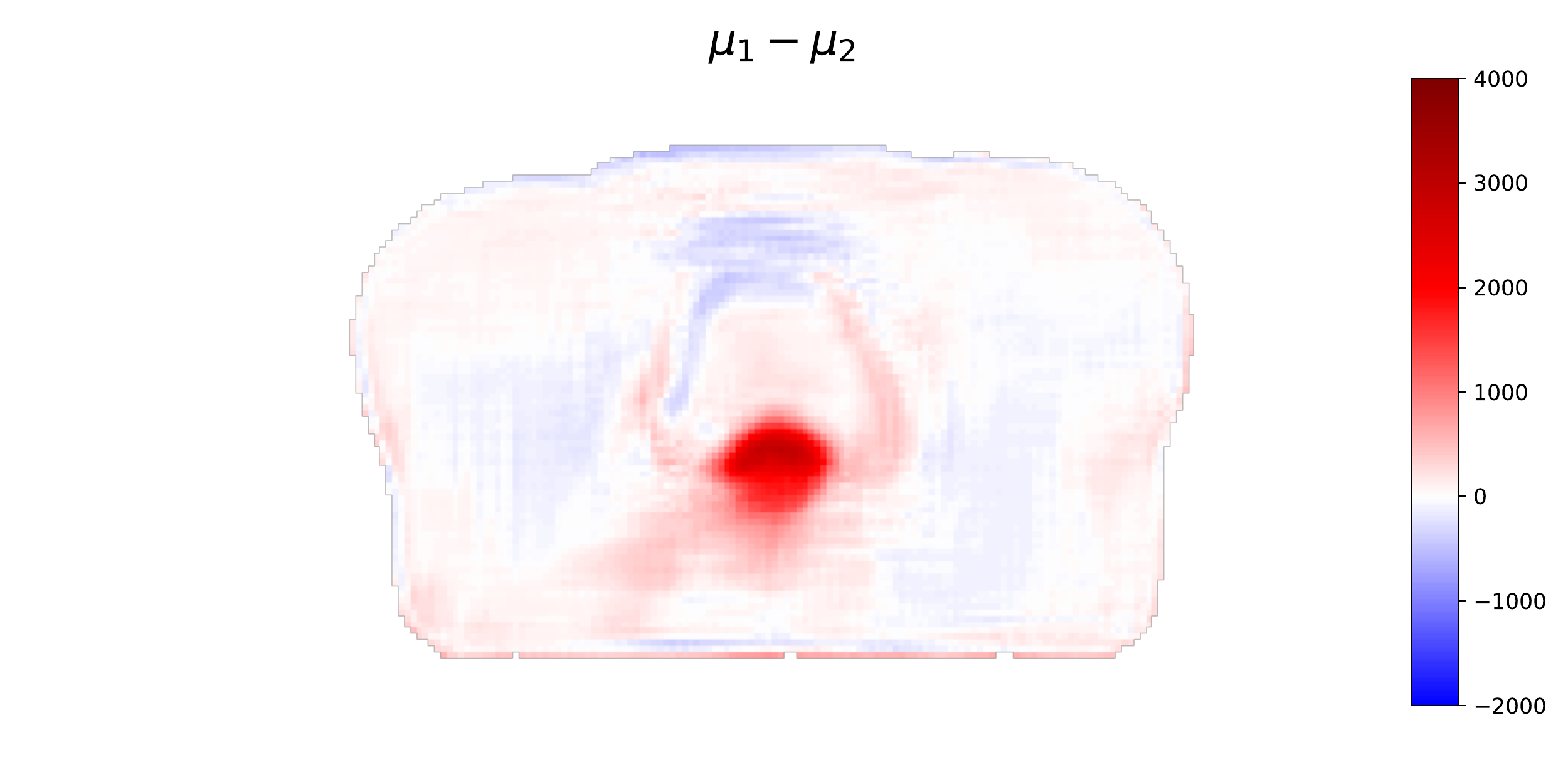} 
    \caption{Dose difference between predicted means $\mu_1$, $\mu_2$.}
    \label{fig:diff-dose}
\end{figure}

\noindent Examining the modes in \cref{fig:doses-and-sigma}, one sees that $\mu_2$ predicts the dimple present in $P_2$, while $\mu_1$ has the the higher target coverage characteristic of $P_1$. Another way to clearly demonstrate that the modes have separated is to plot the difference $\mu_1 - \mu_2$.
In \cref{fig:diff-dose}, one can see that the modes are indeed separated in and around the rectum, while differences in other regions are smaller.

\begin{figure}[h]
    \centering
    \begin{subfigure}{0.7\textwidth}
        \includegraphics[width=1\textwidth]{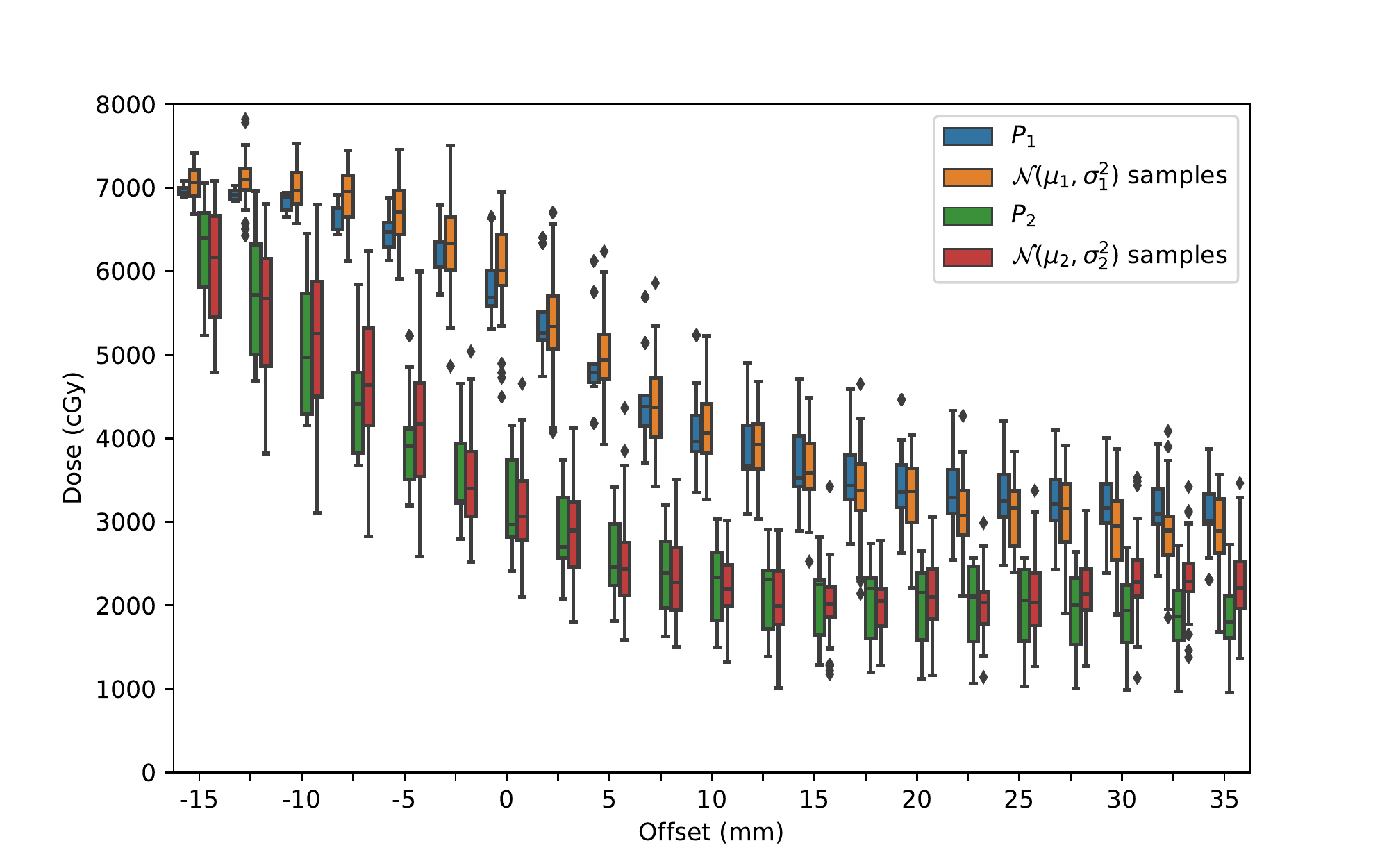}
    \end{subfigure}%
    \begin{subfigure}{0.3\textwidth}
        \includegraphics[width=1\textwidth]{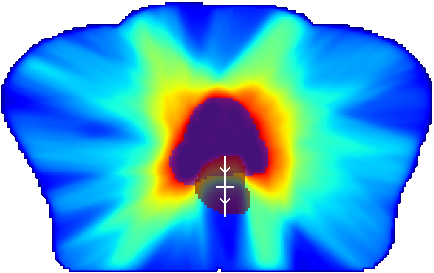}
    \end{subfigure}
    \caption{Distributions of dose values in the predicted mixture components versus the ground truth in the holdout set, on points along a path through the rectum shown to the right. The point at the mass center of the rectum corresponds to an offset of 0. The black small markings show outliers. }
    \label{fig:boxplot-rectum-transversal}
\end{figure}

In order to validate the network across the holdout set (as opposed to only the one patient in figures \ref{fig:doses-and-sigma}, \ref{fig:diff-dose}), we compare the distribution of dose values in equivalent locations for the patients to the predicted counterparts.
The points are drawn along a line in the anterior--posterior direction passing through the center of mass of the rectum in the slice (figures \ref{fig:doses-and-sigma}, \ref{fig:diff-dose}) and corresponding slices through the holdout set.
Samples are drawn from the two predicted components for each patient and compared to the distribution in $P_1$ and $P_2$ using boxplots in \cref{fig:boxplot-rectum-transversal}.
Five samples are drawn for each patient in order to stabilize the empirical distribution and thus the box positions.

\noindent One sees that $\mu_1$ manages to follow $P_1$ well through the drop-off, although a slight discrepancy can be made out when the offset exceeds \SI{10}{\mm}, while $\mu_2$ follows $P_2$ well.

\begin{figure}[h]
    \centering
    \includegraphics[width=0.8\linewidth]{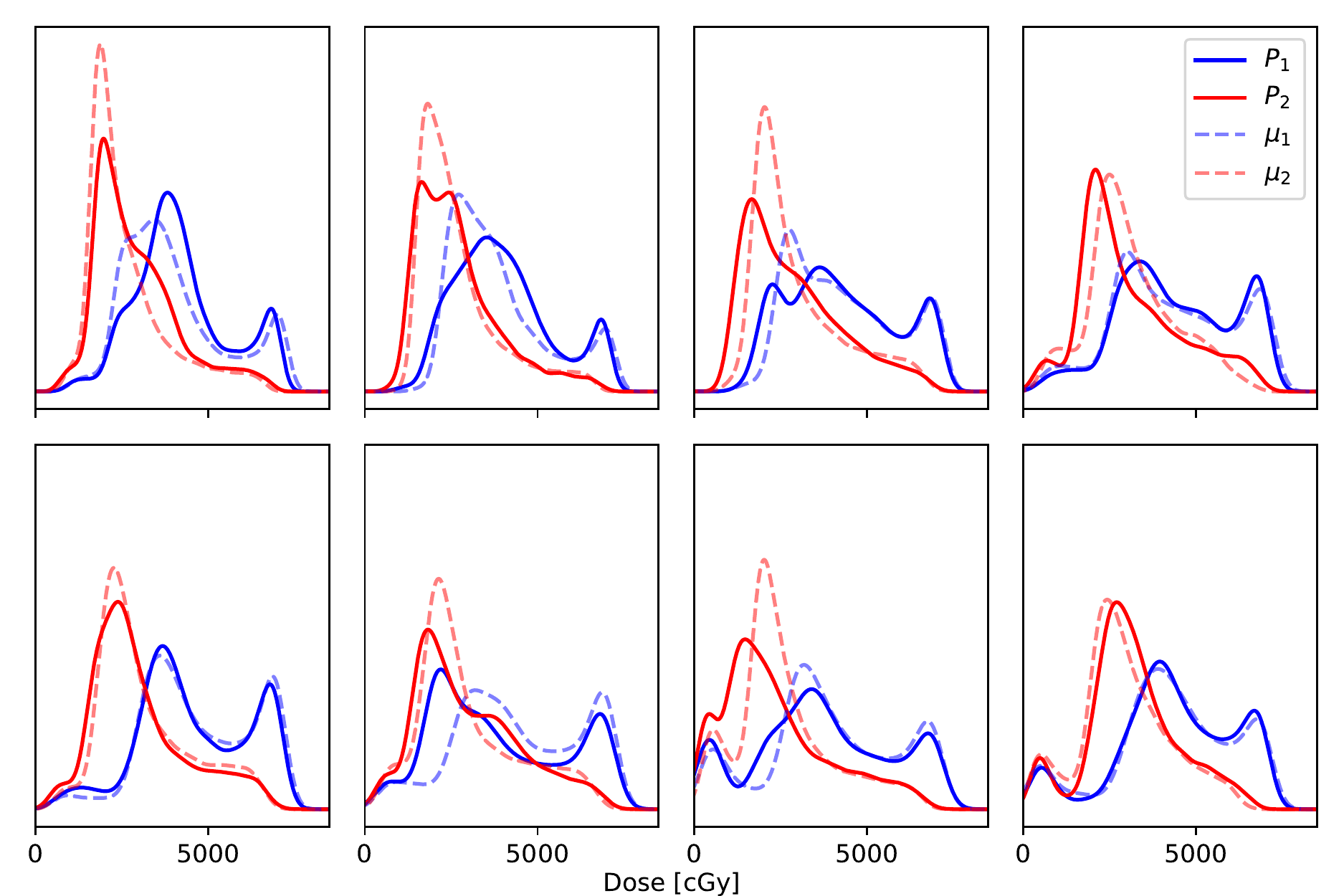}
    \caption{Differental DVHs for the rectum in eight patients. The ground truths $P_1$, $P_2$ are shown next to the predictive modes $\mu_1$, $\mu_2$.}
    \label{fig:kds}
\end{figure}

Illustrations such as figures \ref{fig:doses-and-sigma}, \ref{fig:diff-dose} and \ref{fig:boxplot-rectum-transversal} give a detailed assessment of the model in the most critical area of the patient body, where the PTV is the largest and has the most overlap with the rectum.
As such, however, they focus on one single representative patient and the dose distribution in a vital slice at a time.
A way of examining the behavior throughout the patient body rather than in just one slice, and also for a larger test set, is by comparing differential or cumulative DVHs.
In particular, one would like to see that the DVHs of $\mu_1$ and $\mu_2$ agree with $P_1$ and $P_2$, respectively.
Using kernel density estimates, the differential DVHs for the rectum area for a cohort of test patients are shown in figure \ref{fig:kds}.

\begin{figure}[h]
    \centering
    \includegraphics[width=1.0\linewidth]{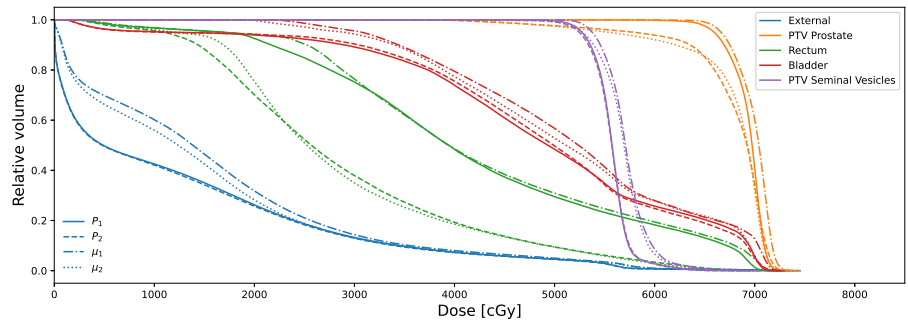} 
    \caption{Average cumulative DVH for the patients in the test set for ground truths $P_1$, $P_2$ and predictive modes $\mu_1$,  $\mu_2$.}
    \label{fig:dvh-testpat}
\end{figure}

In \cref{fig:kds}, it is apparent that the two plans $P_1$ and $P_2$ have created two separate and characteristic distributions in the rectum area. Furthermore, it can be seen that that the predictive modes $\mu_1$ or $\mu_2$ have successfully recreated this behavior, thus reflecting the clinical tradeoff in the region.  
The corresponding cumulative DVHs are also examined for ROIs included in the model.
In \cref{fig:dvh-testpat}, the averages of the DVHs of the predictive modes $\mu_1$, $\mu_2$ throughout the holdout set are shown alongside the ground truth counterparts $P_1$, $P_2$, respectively.
The predicted modes follow the ground truth well for the risk organs and the prostate, essentially capturing the distinct behaviors of $P_1$, $P_2$. 
However, the dose to the seminal vesicles is somewhat overestimated, also causing the predicted DVH of the external ROI to deviate slightly. 
This is not entirely unexpected, considering the fact that the MDN is not trained to necessarily produce predictive means close to the ground truth in terms of DVH.
Apart from the seminal vesicles, the overall plan consistency is kept by the MDN and not only the variations in the rectum are reflected in the predictions.

\begin{figure}[h]
    \centering
    \includegraphics[width=0.9\linewidth]{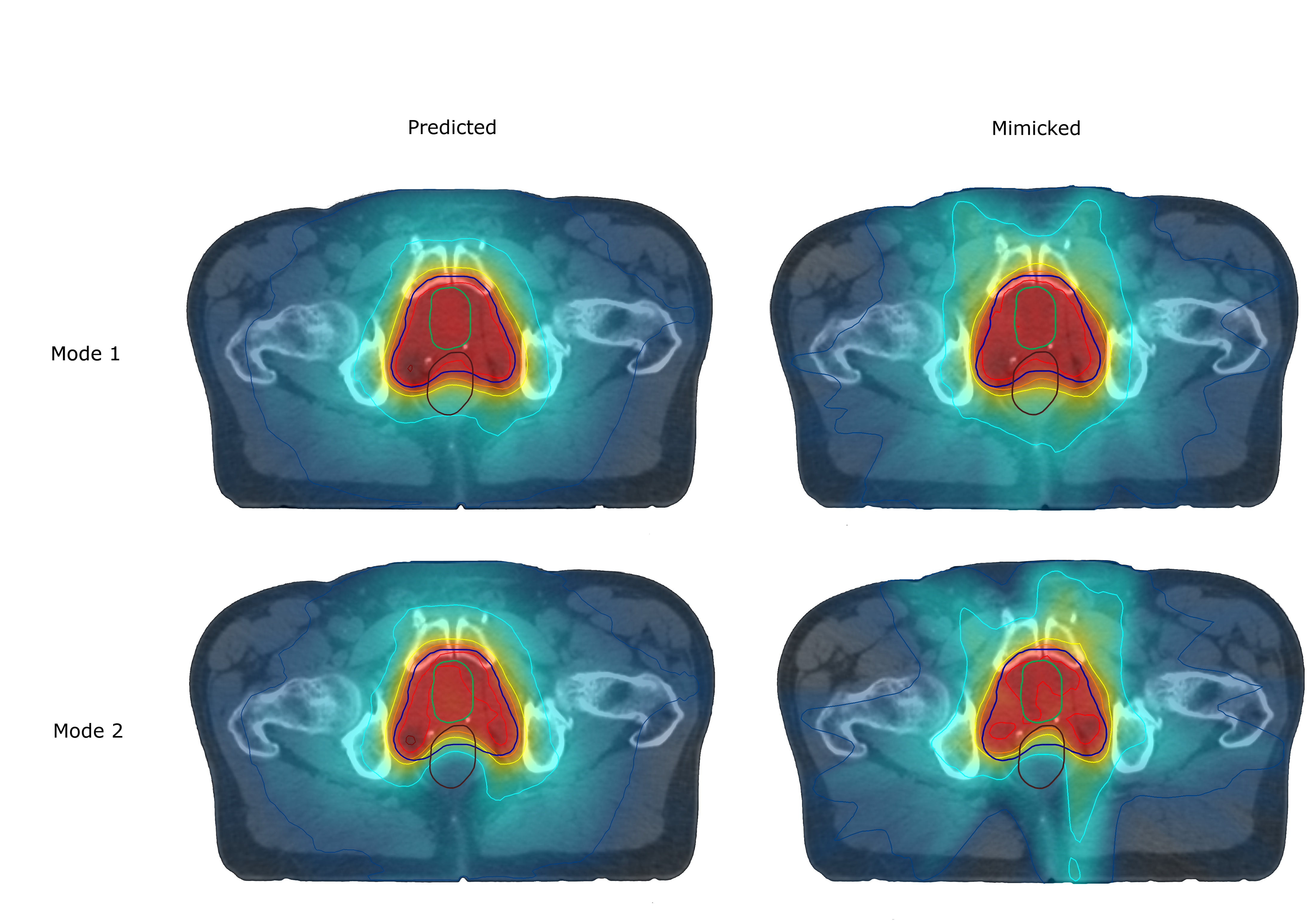} 
    \caption{Comparison of predicted and mimicked spatial dose distributions in a transversal slice for the cases of mimicking $\mu_1$ and $\mu_2$, respectively.}
    \label{fig:mimic_doses}
\end{figure}

\begin{figure}[h]
    \centering
    \includegraphics[width=1.0\linewidth]{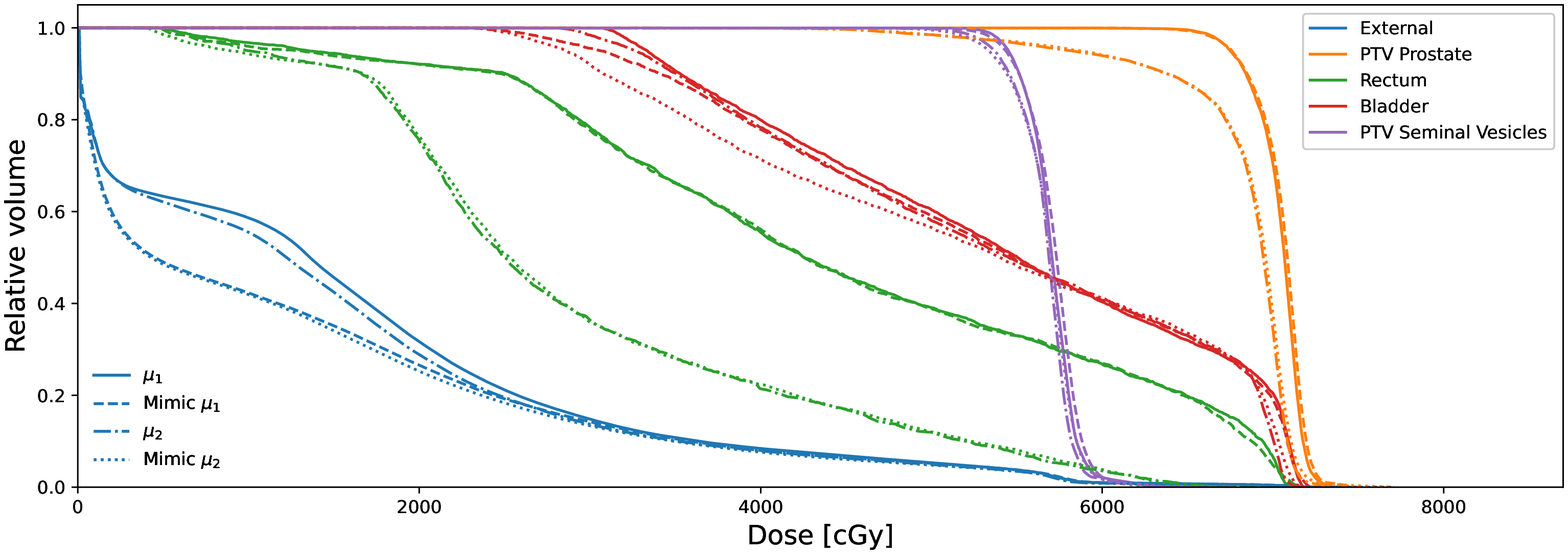} 
    \caption{Comparison of predicted and mimicked DVHs for the cases of mimicking $\mu_1$ and $\mu_2$, respectively.}
    \label{fig:mimic_dvhs}
\end{figure}

Finally, to demonstrate that the proposed dose mimicking method can successfully utilize the predictive distributions outputted by the MDN, we create two deliverable plans for a patient in the holdout set mimicking, respectively, $\mu_1$ and $\mu_2$. Spatial dose and DVH comparisons are shown in figures \ref{fig:mimic_doses} and \ref{fig:mimic_dvhs}. For both cases, one can see that the spatial dose distributions follow the respective predictive modes well, although being slightly less conformal. This is most likely due to the predictions being overly optimistic in the rate by which dose falls off around the target area. One can also see that the DVHs of the mimicked plans followed closely or were in some cases even better than those predicted. While these results show the merits of a probabilistic dose mimicking method, in a clinical setting, one would likely need to further post-process the obtained plans before approval and delivery.

\section*{Discussion}
In this work, we have demonstrated the application of MDNs for probabilistic dose prediction, which has certain advantages when used in an automated RTTP pipeline. We showed that this type of NN can produce predictive probability distributions capturing the uncertain nature of decision making associated with conflicting clinical tradeoffs. This was exemplified on a dataset comprising postoperative prostate cancer patients in which two plans had been created for each patient---one prioritizing target coverage and one prioritizing rectum sparing. Visual inspection across the holdout set showed that the predicted modes followed the respective ground truths well, and the predictive uncertainties were assured to be reasonable by examining the distribution of ground truth dose values around different locations. This has the potential to bring together a comprehensive overview of viable plan alternatives for a given patient, which can not only act as a time-saving tool but also facilitate easier decision making during the RTTP process. In particular, as an example of the usability of a probabilistic prediction, we used a specially developed dose mimicking method to create deliverable plans for each of the predicted mixture components.

While most of previous work has focused on the prediction of a single dose value per voxel, one can, for example, argue that an empirical distribution can be obtained by sampling from generative models, e.g. GANs \parencite{mahmood2018automated, babier2020knowledge} or variational autoencoders \parencite{kingma2013autoencoding}. However, using such approaches, there is no clear way of obtaining the conditional densities $p(d^v \mid x)$ we sought to model in this paper. Also, in the atlas-based method by \textcite{mcintosh2016voxel}, a set of dose propositions are predicted for each voxel by the voting of weak learners, which may be seen as a sample from $p(d^v \mid x)$. The actual density may then be estimated by e.g. kernel density estimation, but compared to MDNs, this approach requires many more parameters to form the densities, and it is moreover not obvious how one would go about separating the modes as was done for our dose mimicking. In these regards, it is concluded that MDNs present advantages over existing methods.


Examples of future work include studying disjoint sets of patients having different treatment protocols and using more realistic data from actually administered clinical plans. 
It would also be of interest to investigate data with other types of tradeoffs---for example, plans with or without a boost volume inside the PTV, or plans with intentional NTCP variations such as for head-and-neck, where saliva glands are spared to different extent in various protocols. 
Depending on the training data, one could look at other types of MDNs, e.g. an NN with more Gaussian mixture components, should the tradeoffs need more modes to be described. 
Future work could also focus on implementation parts, such making the MDN a three-dimensional NN instead of predicting per slice, and various ways to counteract overfitting or stabilizing the training process for this high-dimensional problem that can be prone to mode collapse. 
Another possible future investigation would be to use the knowledge captured by the MDN in multicriteria optimization, which is a natural application of the multimodal predictive distributions where clinically conflicting tradeoffs may be reflected.
 
 
\section*{Conclusion}
The application of MDNs to dose prediction is a promising method toward automation of the RTTP process. 
In this work, we have shown that an MDN can predict dose distributions well and, as opposed to deterministic methods, capture different decision alternatives associated with conflicting clinical tradeoffs. 
In particular, a two-component Gaussian MDN was trained on a dataset of treatment plans for prostate cancer patients with two distinguished degrees of priority of rectum dose sparing over target coverage.
Evaluation on the holdout set showed that the produced probability distributions had predictive means following closely the respective ground truths in terms of spatial dose and partly in DVH, as well as reasonable predictive variances. 
Moreover, deliverable plans were created using a specially developed dose mimicking procedure, exemplifying the usability of such probabilistic information.
In conclusion, the proposed method may serve to assist clinicians in decision making and improve the quality of treatment plans produced by an automated RTTP pipeline.





\section*{Acknowledgments}
The authors thank Adnan Hossain and Mats Holmström for their diligent work on the model used for creating the experimental data, and Tatjana Pavlenko, Pierre Nyquist, Fredrik Löfman and Marcus Nordström for valuable input on the manuscript. 



\printbibliography

\end{document}